# Effect of ultrasonic irradiation power on sonochemical synthesis of gold nanoparticles

J.A. Fuentes-García[1,2], J. Santoyo-Salzar[3], E. Rangel-Cortes[4], G.F. Goya[1,5], V. Cardozo-Mata[4] and J.A. Pescador-Rojas[4*]

[1] Instituto de Nanociencia de Aragón (INA), Universidad de Zaragoza, 50018 Zaragoza, Spain.
[2] Unidad Profesional Interdisciplinaria en Ingeniería y Tecnologías Avanzadas del Instituto Politécnico Nacional, UPIITA-IPN, Av. IPN 2580, Ticoman 07340 Mexico.
[3] Departamento de Física, Centro de Investigación y de Estudios Avanzados del Instituto Politécnico Nacional, CINVESTAV-IPN, Av. IPN 2508, Zacatenco 07360, México.
[4] Universidad Autónoma del Estado de Hidalgo. Escuela Superior de Apan. Carretera Apan−Calpulalpan Km.8, Col. Chimalpa, 43920 Apan, Hgo., México.
[5] Departamento de Física de la Materia Condensada, Facultad de Ciencias, Universidad de Zaragoza, 50009 Zaragoza, Spain.
∗Corresponding author: nanopoli@gmail.com

**Abstract**

In this work, optimized size distribution and optical properties in colloidal synthesis of gold nanoparticles (GNPs) were obtained using a proposed ultrasonic irradiation assisted Tuerkevich-Frens method. The effect of three nominal ultrasound (20kHz) irradiation powers: 60, 150 and 210 W has been analyzed as size and shape control parameter. The GNPs colloidal solutions were obtained from chloroauric acid ($HAuCl_4$) and trisodium citrate ($C_6H_5Na_3O_7 \bullet 2H_2O$) under continuous irradiation for 1 hour without any additional heat or stirring. The surface plasmon resonance (SPR) was monitored in the UV-Vis spectra every 10 minutes to found the optimal time for localized SPR wavelength ($\lambda_{LSPR}$) and the 210 sample procedure reduces the $\lambda_{LSPR}$ localization to 20 minutes, while 150 and 60 samples show $\lambda_{LSPR}$ in 60 minutes. The nucleation and growth of GNPs showed changes in shape and size distribution, which were associated with physical (cavitation, temperature) and chemical (radical generation, pH) conditions in the aqueous solution. The results showed quasi–spherical GNPs as pentakis dodecahedron ($\lambda_{LSPR}$=560 nm), triakis icosahedron ($\lambda_{LSPR}$=535 nm), and tetrakis hexahedron ($\lambda_{LSPR}$= 525 nm) in a size range from 12-16 nm. Chemical effects of ultrasound irradiation were suggested in the disproportionation process, electrons of $AuCl_2^-$ are rapidly exchanged through the gold surface. After $AuCl_4^-$ and $Cl^-$ are desorbed and a complex tetrachloroaurate is recycled for the two-electron reduction by citrate, aurophilic interaction between complexes $AuCl_2^-$, electrons exchange and gold seeds, the deposition of new gold atoms on the surface promoting the growth of GNPs. These mechanisms are enhanced by the extraordinary effects of ultrasound such as the cavitation effects and transmitted energy into the solution. These results show the that the plasmonic response from our nanoparticles can be tuned using a simple methodology with minimum infrastructure requirements. Moreover, the production method could be easily scalable to meet industrial manufacture needs.

**Keywords:** Sonochemistry, gold nanoparticles, irradiation power.

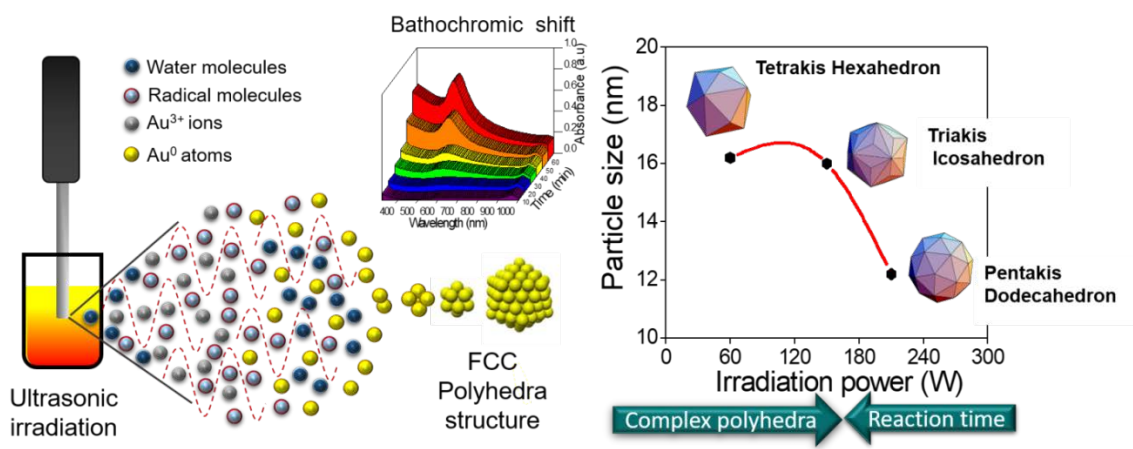

## 1. Introduction

The gold nanoparticles (GNPs) have been studied as plasmonic materials for their application in biomedical field. Some of these applications involves photo-thermal cancer treatment agents [1], molecular labeling [2], radiotherapies [3], drug delivery systems [4], and optical sensing devices [5]. The current demand for well-defined GNPs properties is creating the cutting edge methods toward more efficient, safe and economically for nanoparticle manufacture [6]. Colloidal GNPs have been easily synthesized and size and shape controlled by chemical methods [7]. Turkevich *et al.* developed a synthetic route to obtain colloidal gold nanoparticles (GNPs) based on hydrogen tetrachloroaurate ($HAuCl_4$) reduction from citric acid and sodium citrate as reducing agents in boiling water, under vigorous stirring conditions [8]. The advantages of sodium citrate salt in the synthesis are: it acts as a reducing and stabilizing agent, also forms a buffer in aqueous solution avoiding abrupt pH modifications in the reaction. Tuerkevich-Frens method allows preparing GNPs with different sizes by modifying gold ions/citrate ratios for controlled nucleation velocity and steric surface can be achieved for colloidal stability and biological functionalization [9-11].

A challenge in the chemical synthetic routes for GNPs manufacture is to control the primary morphology, particle size distribution, crystalline structure in a simple way, and retain their properties along time [12]. Several strategies have done to control these properties using surfactants [13], dispersing agents [14], and biological compounds [15]. Likewise, other approaches include electromagnetic irradiation, such as microwaves [16], ultraviolet (UV) light [17], electron beam [18], and gamma-ray [19] to obtain GNPs without using reducing agents, and their sterilization by means of a one-step process were also developed [20]. The interaction between irradiation energy and the irradiated solution has been the key to control the properties of GNPs by kinetics reaction manipulation. However, high energy is needed for ionizing those systems, additionally, complicated experimental setups are required, and making manufacture non-rentable for industrial production.

Ultrasound processing is an environment-friendly, safe and inexpensive technology [21], used as a versatile tool in a wide range of scientific and technological applications, for instance: biology [22], chemistry [23], physics [24], medicine [25], material science [26], and industrial applications [27-30]. Nanomaterials can be obtained and modified using ultrasonic irradiation to assist synthetic methods in aqueous solution [31]. The acoustic cavitation produced by ultrasonic irradiation in water generates transient bubbles, within which, different physical and chemical phenomena generated by their implosion play an important role in the formation of nanostructures [32].

Ultrasound irradiation can produce the conditions for chemical agent-free reduction of gold ions [33] and noble metal nanoparticles formation in different morphologies [34]. Combining the ultrasound physical effects such as: cavitation, arises of temperature and local pressure differentials (primary sonochemistry) with chemical reactions promoted by hydrogen and hydroxyl radicals produced by sonolysis (Eq. 1) (secondary sonochemistry).

$H_2O$ + ultrasound → $e_{aq}^-$, $H_3O^+$, $H^\bullet$, $H_2$, $OH^\bullet$, $H_2O_2$ 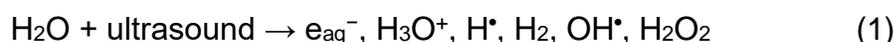 (1)

The bubbles volume and life cycle are determined by ultrasonic power intensity in aqueous media [34]. From its formation, growth, and final implosion as result of their collapse, they induce hot spots production with internal local temperatures up to thousands of K and high pressure differentials [35]. At these conditions, the reactants have intensive interactions in extremely short time-lapse, cavitation events results in heating and cooling rates of more than $10^{10}$ K s$^{-1}$ and sonocrystallization can be improved [36]. Secondary sonochemistry effects are dominant at 20–100 kHz ultrasound frequencies with a higher level of transient cavitation, whereas chemical effects are dominant in range of 200–500 kHz due to the generation of a large number of active bubbles [37].

Scientific reports have analyzed the effects of the sonochemical parameters in GNPs synthesis; i.e. frequency [38], low irradiation power (<100 W), and stabilizing compounds concentration [39]. However, ultrasonic irradiation power is the main factor of chemical intermediates generation at pressure−temperature relationships during the oxidation−reduction reaction [40].Then, investigate their effect in GNPs is important to enhance their optical properties for Biomedical applications based on intrinsic features such as SPR, physicochemical and surface interactions to recognize biomolecules in sensing systems or photo thermal and photodynamic therapies [41]. The GNPs properties are sensitive to size and shape modifications and their adequate control during the synthesis is crucial for their implementation as tool in early disease detection [42] and multifunctional nanomedicines for cancer therapy [43].

Furthermore, the molecular dynamics simulations suggest that the surface energy of GNPs decreases by increasing cluster size at 0 K, but it is increased at higher temperatures [44]. The self-assembly of molecular polyhedra into complex structures can modify the shape of $Au^0$ atoms polyhedral in a face-centered cubic (FCC). The FCC clusters have shown different optical behavior depending on the facet structure, the crystal system arrangement is crucial in the modification of ultraviolet and visible (UV-Vis) interactions. Au dodecahedron and icosahedron exhibit bathochromic shifted and broader surface plasmon resonance (SPR) peaks, whereas the quasi−spherical shape of the truncated octahedron leads to a similar optical response as a perfect sphere, with a slight bathochromic shifting and localized surface plasmon resonance (LSPR) [45]. Electron diffraction patterns observed in transmission electron microscopy (TEM) characterization have confirmed the facet orientations and the FCC structure to form exotic solids [46]. An experimental survey on the Au clusters synthesis reported their evolution in different solids, controlling the cluster arrangement is possible to influence the optical response [47].

In this work, the variation of ultrasonic power from 60 to 210 W nominal values in the synthesis of GNPs using an ultrasonic probe at 20 kHz of frequency was studied. The experimental methodology was designed to observe the primary and secondary sonochemical effects monitoring temperature every 10 minutes and final pH differential after 60 minutes of irradiation to determine the macroscopic effects of cavitation and radical formation in the aqueous solution volume, which acting together with sodium citrate for improved reduction of chloroauric acid. The SPR formation and localization inspected by UV-Vis band around 520 to 560 nm of 10 minutes aliquots confirms the reduction of $Au^{+3}$ to $Au^0$ ions and the subsequent Au clusters formation. The obtained GNPs were characterized by transmission electron microscopy (TEM) to observe their size, shape and Fast Fourier Transform (FFT) of high-resolution images revealed their FCC structure. These results show the influence of ultrasound in the GNPs production, increasing irradiation power as a strategy to controlled size, shape, and SPR in a simple way and optimized time.

## 2. Materials and methods

*2.1 Materials and equipment*

Reagents used were Tetrachloroauric(III) acid ($HAuCl_4$ ≥ 99%, Sigma Aldrich), Trisodium citrate dihydrate ($C_6H_5Na_3O_7 \bullet 2H_2O$, ≥ 99%, Sigma Aldrich) and deionized water (18MΩ, ≤ 4.3 µS/cm, Millipore). Ultrasonic Homogenizer 300VT BioLogics, Inc. was used with titanium sonotrode (diameter=9.5 mm, length=108 mm). Temperature increasing was detected using an immersed thermocouple (K Type) in the irradiated solutions. The pH values were measured using HANNA Instruments, HI-2210-02 Bench Top pH Meter with glass electrode. UV−Vis spectra were obtained from a Thermo Scientific UV−Vis GENESYS 10S spectrophotometer. TEM observation and HRTEM images were performed in an image−corrected FEI Titan$^3$ at 300 kV. FFT images were obtained by using Gatan Digital Micrograph software.

*2.2 Sonochemical synthesis of gold nanoparticles*

Following the Turkevich−Frens method [9], 50 mL of chloroauric acid ($HAuCl_4$), 0.025 mM was poured into a 100 mL beaker (Borosilicate glass), which was added 1 mL of 1.5% (w/v) aqueous solution of trisodium citrate ($Na_3Ct$) under ultrasonic irradiation at 60, 150 and 210 W for one hour, at room temperature. The $Na_3Ct/HAuCl_4$ ratio used in the three samples was 3:1 (w/v). Samples were labeled as 60, 150, and 210 respectively. A schematic representation of the experimental ultrasonic processing for GNPs, and the sonochemical effects in aqueous solution are showed in Fig. 1.

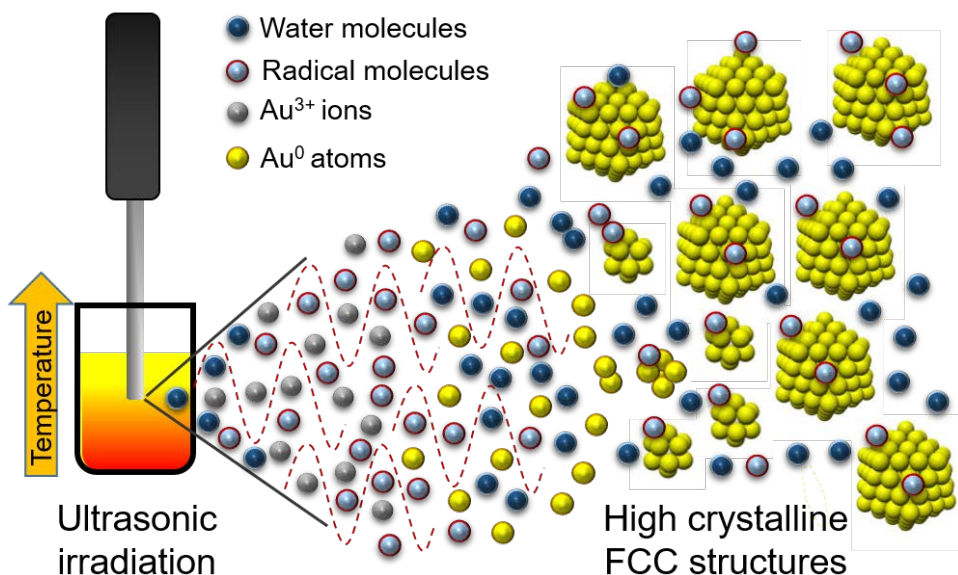

**Fig. 1.** Processing of GNPs, shows nucleation and growth promoted by sonochemical effect in ultrasonic irradiation in aqueous solution.

*2.3 Characterization*

The UV−Vis spectra (350 to 1000 nm) were collected from 3 mL aliquots every 10 minutes and measured in a quartz cuvette for monitoring the SRP bathochromic shift during GNPs nucleation and growth. Samples were supported on Lacey copper grids by dropping a diluted sample in water and drying in a vacuum for 24 hours for TEM observation. During the colloid synthesis temperature readings every 10 minutes were made for all experiments. The pH was measured after and before of ultrasonic irradiation.

## 3. Results and Discussion

*3.1 Optical characterization*

After performing the ultrasound radiation, the colloidal solutions showed different colors, violet for 60 W and ruby-red for 150 and 210 W samples. Figs. 2a−c shows the UV−Vis bathochromic shifting observed in the spectra, as a function of time reaction for all synthesized samples. The GNPs sample at 60 W had a limited intensity from absorption due to nucleation and growth, this effect is associated with large Au clusters formation, but does not contribute to the absorption band signal. The spectra revealed a broadband localized at 650 nm after 60 minutes of irradiation (Fig. 2d). The hypochromic shift at $\lambda_{SRP}$= 520 nm was associated with small GNPs signal and the highest occupied molecular orbital (HOMO). The lowest unoccupied molecular orbital (LUMO) value of 2.21 eV probes the formation of large aggregates. The broad full width at half maximum (FWHM) (114 nm) of the SRP band was associated with the multiple twinned GNPs in FCC orientation and aggregate formation.

The LSPR band obtained in the UV-Vis spectra of 150 and 210 samples probe optical properties modification in reduced time as ultrasound power increases. Fig. 2b shows a narrow LSPR band ($\lambda_{LSPR}$ = 535 nm) formation in UV-Vis spectrum of 150 sample, it takes 60 minutes of irradiation. Whereas, the 210 sample (Fig. 2c, $\lambda_{LSRP}$= 525 nm) occurs after 20 minutes (Figure 2f). The FWHM reduction from 82 to 75 nm (Fig. 2e) and the increase in HOMO-LUMO energy values from 2.32 to 2.36 eV (see Table 1) suggested quantum confinement effects related to reduced size and monodispersion of as obtained GNPs. Is possible to manufacture GNPs with high absorption interaction using ultrasonic irradiation as an only energy source, reducing laboratory requirements and controlling properties modifying simple parameters.

According with literature [48], narrow SRP bands (reduced FWHM values) correspond to interactions of monodisperse particles with UV-Vis radiation and the bathochromic shifting was related to complex polyhedral arrangement of $Au^0$ atoms in GNPs FCC structure predicted in theoretical simulations. The proposed ultrasound assisted Turkevich-Frens method allows to observe these spectral changes. During the experimental procedure, UV-Vis interactions reveals monodisperse and high LSPR GNPs in a few minutes (Fig. 2e). The results can be associated to an optical enhancement thanks to their accurate response in a specific interval. It can be interesting for biomedical applications based in the GNPs radiation absorption such as photo thermal therapies.

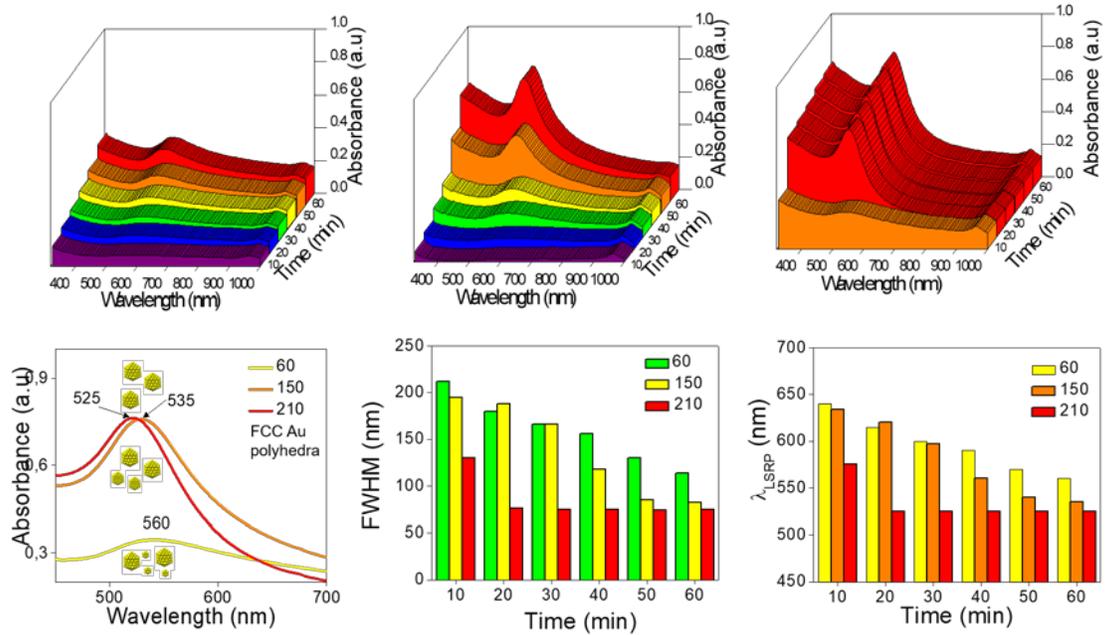

**Fig. 2.** GNPs growth dynamics at different irradiation powers (a) 60 W, (b) 150 W, and (c) 210 W monitoring surface plasmon resonance in UV-Vis spectra, showing the relationship among absorbance, wavelength and synthesis time. (d) shows the proposed evolution of dispersion and size of GNPs face-centered cubic (FCC), (e) summarizes the FWHM evolution of SRP bands, narrowing can be achieve in 20 minutes at 210 W, and (f) LSPR and bathochromic shift shows the influence of ultrasound power increasing in the optical properties of GNPs.

*3.2 Structural characterization*

Representative TEM images of GNPs samples are shown in Fig. 3.The particle size distribution of 60 W sample was 16±3 nm. In this case, large aggregates and isolated nanoparticles were visible, matching with UV−Vis spectra predictions. The GNPs crystalline structure was defined by HRTEM as faceted tetrakis hexahedron polyhedral arrangement (Fig. 3b), while the selected area electron diffraction (SAED) pattern (Fig. 3c) confirms the complex array. The 150 sample (Fig. 3d) reveals triakis icosahedron crystallization and reduced polydispersion (16±1.6 nm).

However, particle size have not perceptible changes (Fig. 3e). In the 210 sample, the size was reduced to 12±1 nm (Fig. 3g). A quasi-spherical morphology was observed as faceted pentakis dodecahedron (Fig. 3h) confirmed by the rings in the simulated electron diffraction patterns image (Fig. 3i).

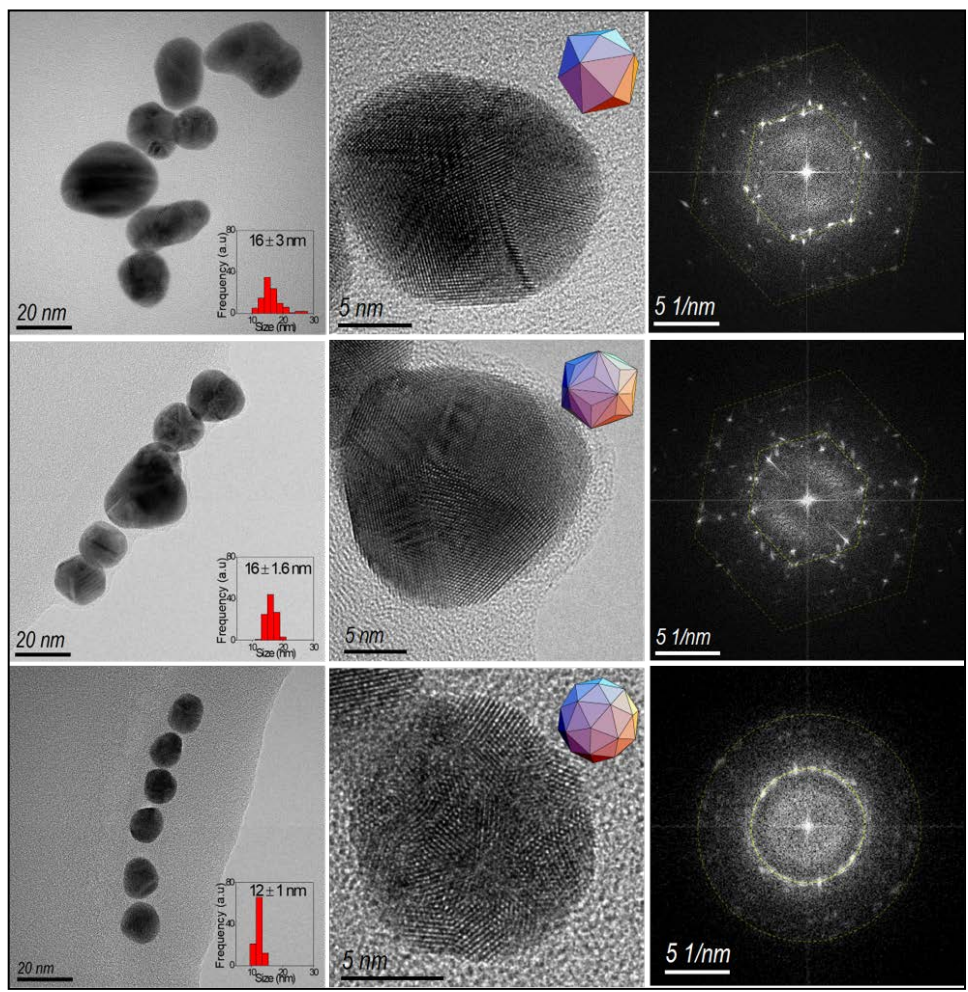

**Fig. 3.** In the first column, the (a), (d), and (g) TEM micrographs of nanoparticles, in the insets, histograms of size distribution are the showed. In the second column (b), (e) and (h) are HRTEM micrographs, insets are polyhedral. The third column (c), (f), and (i) allows us to observe the SAEDs.

Smaller sizes and more spherical clusters were produced by increasing irradiation power, according with the structural characterization. The final size depends on the relationship between the metallic precursors concentrations and the number seeds formed under the synthesis conditions, as previous works have been reported. If there are high concentration seeds in the system, the final particle size will be small [11]. Under the proposed experimental conditions in this work, smaller nanoparticles were obtained increasing nominal irradiation power to a maximum value of 210 W at constant precursor concentration for the three samples. The size reduction suggest that the concentration of seeds is higher compared with other two samples (60 and 150) and ultrasound power increase improves the seed rate in solution. Temperature arise rate produced by primary sonochemical effects (Fig. 4a) increase the particle size and dispersion as described by Wuithschick et al. [49], and Takiyama [50]. They have observed GNPs size and polydispersion increase as temperature during the synthesis using borohydride and citrate as reductors.

However, the three samples reported in this work, have different heat rates and after 20 minutes of temperature arise, it reaches a constant temperature value. Therefore, the influence of primary and secondary sonochemical effects in GNPs synthesis can be a combination of processes [51]. Increasing the thermal energy ($E_{kT}$) on the reaction system, a change in equilibrium towards a more hydroxylated form ($HAuCl_4$ solution) can be promoted. Decreasing the redox potential of $[AuCl_4]^-$ as pH increases the size of the GNPs obtained is dominated by the number of nuclei formed [11]. At this point we can associate the heat rate reached using ultrasonic irradiation with the velocity of nucleation and seed formation in GNPs synthesis and the stable region of Fig.4a with the growth time. It is supported by the reduced time of LSPR in 210 sample (Fig.2c). After 20 minutes any changes in SPR wavelength were detected, it suggest the seeds are exhausted in the colloid and monodisperse GNPS can be achieved. The 60 sample shows a broad SPR band manifesting strong contributions of polydisperse GNPs. The reduced velocity of nucleation reached using 60 W limits the seed formation and bigger GNPs were obtained.

**Table 1** Experimental data summarized from left to right: nominal irradiation power (W), heat rise rate (°C/min), gradient of final and initial temperature (room temperature), average size of GNPs and respective standard deviation, localized surface plasmon resonance (LSPR) wavelength (nm), full width at half maximum (FWHM) values of UV-Vis

spectra after 60 minutes of ultrasonic irradiation (nm), energy difference between the Highest Occupied Molecular Orbital (HOMO) and Lowest Unoccupied Molecular Orbital (LUMO) or HOMO–LUMO gap (eV), final pH values, and in the last column, the different types of obtained GNPs polyhedral configuration.

| Irradiation power (W) | Heat (°C/min) | ΔT (°C) | Average Size (nm) | $\lambda_{LSPR}$ (nm) | FWHM (nm) | HOMO-LUMO (eV) | $pH_{final}$ | Polyhedra |
|---|---|---|---|---|---|---|---|---|
| 60 | 2 | 34 | 16.2±3 | 560 | 114 | 2.21 | 5.09 | Tetrakis hexahedron |
| 150 | 3 | 42 | 16±1.6 | 535 | 82 | 2.32 | 5.23 | Triakis icosahedron |
| 210 | 4.5 | 52 | 12.2±1.2 | 525 | 75 | 2.36 | 5.31 | Pentakis dodecahedron |

The sonochemical method implemented in this work, did not reach boiling temperature unlike the classic hotplate system. However, the latent heat in the reaction medium causes a small but notable reduction in the size of GNPs obtained in the citrate reduction of $HAuCl_4$ in water. Cavitation promotes local violent reaction conditions, heating and turbulent agitation by heat transfer of gas bubbles [52], driving to accelerate nucleation process and concentration of seeds could be elevated. These phenomena lead to the relaxation and interface defects formation, in order to form polyhedral GNPs. Likewise, the energy is absorbed by the system during the cavitation processing into the solution. Then this can lead to the exotic polyhedral morphologies formation, as have been predicted in the literature. High energy potentials and formation systems can improve the plasmonic responses [53] (Fig. 4b).

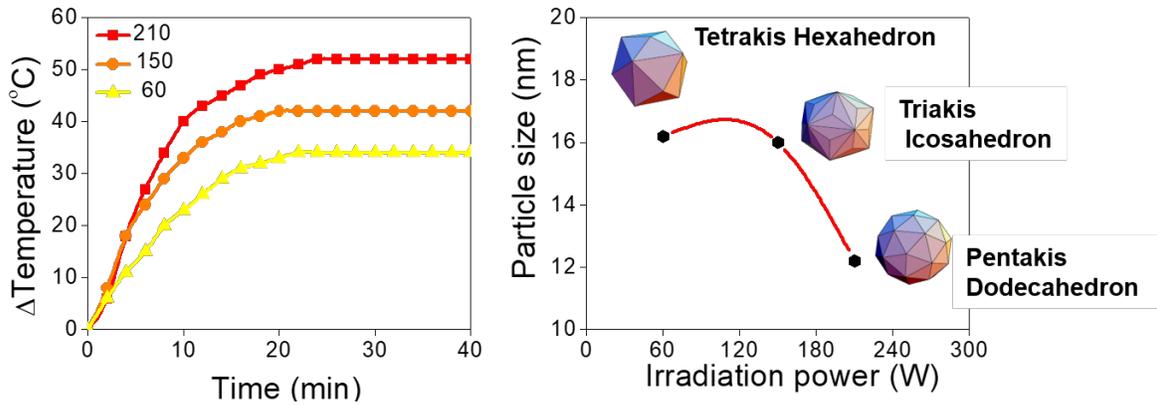

**Fig. 4.** Left panel: Temperature increase as a function of time along the reaction process for samples labeled 60, 150 and 210. The stabilization of the reaction temperature was observed after approximately 20 minutes from the start of the irradiation. Right panel: illustration of the changes in polyhedron shapes adopted by GNPs and the corresponding average particle sizes *vs.* nominal irradiation power.

Energy transmitted to the solution can be expressed as ultrasound power (W), ultrasound intensity (W/cm$^2$), acoustic energy density (W/mL), or cavitation intensity. The ultrasound and the cavitation activity in a reactor may vary for the same ultrasound intensity, if the sample volume and location of ultrasound transducer changes. Ultrasonic power, intensity, and acoustic energy density can be calculated using the following equations [37]:

$$Power\ (W) = mC_p \left[\frac{dT}{dt}\right]_{t=0} \qquad (2)$$

$$Ultrasonic\ intensity\ (W/cm^2) = \frac{P}{A} \qquad (3)$$

$$Acoustic\ energy\ density\ (W/mL) = \frac{P}{V} \qquad (4)$$

where, $m$ is the mass ($0.09\ kg$), $C_p$ is the specific heat capacity ($4.186\ \frac{kJ}{kg\ K}$), $A$ is the area of the radiating surface (in the case of ultrasonic probe $A = \frac{\pi D^2}{4}$ where $D$ is the diameter ($0.95\ cm$), $V$ is the volume ($90\ mL$), and $(\frac{dT}{dt})$ is the initial rate of change of temperature during sonication. According with [37], this can be determined by fitting the data of Figure 4 (a) to a polynomial curve and extrapolating to time $(t) = 0$. In Table 2, calculated power in function of the nominal irradiation power. The transmitted energy in the system increases, then raised disruption caused by ultrasonic intensity reduces the lifetime of transient bubbles and cavitation effects are more evident. These effects allow tuning the energy supplied to the system during the synthesis in order to control GNPs properties. The obtained values of the primary sonochemical effects in the solution are similar and lower than the irradiation conditions used in food industry applications [54].

**Table 2.** Values of nominal irradiation power and experimental irradiation power, ultrasonic intensity, and acoustic energy density determined using calorimetric method.

| Nominal irradiation power (W) | Experimental irradiation power (W) | Ultrasonic Intensity (W/cm$^2$) | Acoustic energy density (W/mL) |
|---|---|---|---|
| 60 | 0.76 | 1 | 0.0111 |
| 150 | 1.12 | 1.5 | 0.0124 |
| 210 | 1.58 | 2.2 | 0.0175 |

The chemical effects in the proposed GNPs ultrasound assisted synthesis can be explained as follows. During the reduction there will be at least four different anions present in the reaction medium: gold(III) chloride; ($AuCl_4^-$); citrate; chloride; and hydroxide anions [55]. Following mechanism outlines, an important factor is competitive and preferential adsorption of $AuCl_4^-$ ions instead of citrate ions on the surface of formed seeds [56]. The $AuCl_4^-$ cannot to be adsorbed as a charged ion: either counter-ion complexation occurs or one of the $Cl^-$ ligands is displaced when adsorption takes place, effectively adsorbing as $AuCl_3$ [57]. To the previous mechanism can be added that, effective absorption of ultrasonic energy can be achieved if GNPs are in resonance with the ultrasound. Furthermore surface potential of the pure gold surface [58], becomes more negative with increasing pH [55]. Then, is possible improve the control of the homogeneity, size and shape of the synthesized nanoparticles through the combination of ultrasound and citrate in aqueous medium without the need to use mechanical stirring and external heating [38]. The presented results in this work probe that it can be achieved modifying only ultrasonic power input

In Scheme 1, reaction mechanism for the reduction of chloroaurate is illustrated. Ojea-Jiménez and Campanera [59] have stated that during the initial stage, the chloride ion must be displaced from chloroaurate by an oxygen atom in the citrate basic form. The intermediate I undergo an internal reorganization of the hydrogen bridge, which coordinates with gold (I→II). In the later stage, two-electron are transferred and decarboxylation occurs, chloride displacement yields 3-oxoglutaric acid (3-OGA) by citrate and aurous chloride complex.

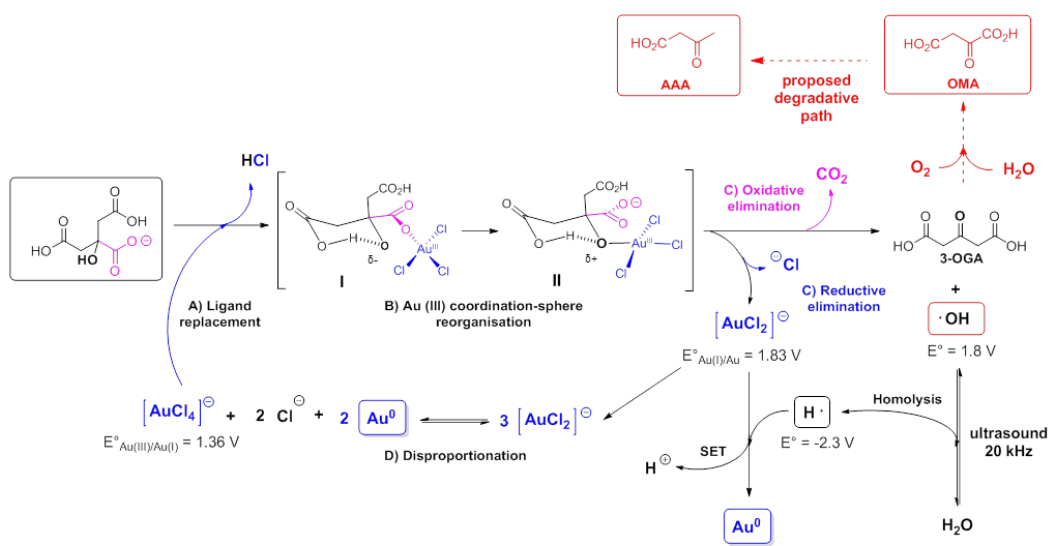

**Scheme 1.** Mechanistic proposal presenting combined chemical and ultrasonic reduction of chloroaurate. A) Ligand replacement: negatively charged oxygen atom from carboxylate group in citric acid, substitutes a chloride ligand of tetrachloroaurate complex anion (with formal loss of a molecule of chlorhydric acid equivalent). B) Au (III) coordination-sphere reorganisation (adduct I → II): it is a two-step representation of necessary assembly of electron arrangement in order to allow the reductive elimination step. C) Reductive elimination: the two electrons constituting O-Au (III) bond will be transferred to core shell of gold atom, passing from +3→+1 formal oxidation state in adduct II. D) Oxidative

elimination: the carbon dioxide loss from adduct II generate 3-oxoglutaric acid as by-product from citric acid. Homolysis: homolytic cleavage of solvent generates local accumulation of high-energy species such as hydroxyl and hydrogen radicals capable of react in the periphery of implossing cavitatory bubbles.

The final reduction stage poses at least three possibilities: According to Ojea-Jiménez and coworkers [60], 3-OGA induces the fast [3-OGA-Au+] complex formation that promotes the stability of precursor by disproportionation of aurous species. It is well-known the ultrasonic energy promotes chemical reactions by pyrolysis (thermal dissociation) water molecules and forms •OH and •H highly reactive free radicals [61].

Another arrangement adequate to disproportion reaction, with no need for 3-OGA intervention, is based on the intensity of aurophilic interactions between gold atoms and/or gold (I) ions with energies comparable to strong hydrogen bonds [59]. We consider the formation of small clusters of mixed metal-ion gold, as has been demonstrated for silver species [63], as understood, 3-OGA is capable of decomposing in the presence of transition metals [64] through a SET mechanism involving the reduction of gold (I) to a gold atom, with the formation of a radical intermediate, once again.

During the disproportionation process, a complex tetrachloroaurate is recycled for the two-electron reduction by citrate. The equilibrium is driven by the departure of gold from the system to form clusters, which eventually yields seeds for NPs growth [65, 66]. These perform the catalyst role for deposition of newly formed gold atoms because of reaction proceeds faster in the presence of gold surfaces, as a seed-mediated growth mechanism [11]. In this case, no further seed formation was observed in the SPR and the initial settled kinetic control became thermodynamic.

A second reduction mechanism is proposed based on the range of hydrogen produced in the reaction medium. At working pH values of 3.3 (initially) and up to 5.2 after the reaction, the formation of reductants hydrated electrons [67] in this solution is not feasible. Thus, we conclude that hydrogen is the main reducer in a mechanism single electron transfer SET. The reduction potential was −2.3 V. It readily reduces inorganic ions, like Au(I) yielding Au(0) and hydronium ion as products, like the reduction of the silver ion, as presented by Sarkar [68]. The degenerate reaction of the hydrogen atom with the hydroxyl radical reforms the solvent. Even when $H_2O_2$ is formed, this has been used to promote nanoparticle growth after seed addition to a chloroaurate solution [69]. The hydrogen ranges produced with this methodology are from 20 to 18 µM, other reported ranges are between 10 and 6 µM per hour (on sonochemical oxidation of iodine by a hydroxyl radical at 20 kHz and 42 ±2 W) at room temperature and 50 °C respectively [70].

The third reaction mechanism is based on the radical species formed in the reaction. The presence of air dissolved in the reaction mixture triggers a radical chain mechanism that involves oxidative degradation of citric acid [71]. Notably, the oxidative pathway of citrate, 3-oxoglutarate and further, encompasses many propagation steps, including possible formation of an alkyl radical, centered on the transfer of a carbon atom $\alpha$ to the carboxyl group. NMR has detected oxidative byproducts during the formation of gold nanoparticles by citrate reduction [72]. The oxidation implies the formation of a carbocation highly unstable, by the electron-withdrawing group (EWG) –$CO_2H$. The third reaction mechanism is based on the radical species formed in the reaction EWG [73].

## 4. Conclusions

The proposed experimental methodology using ultrasound to GNPs manufacture allowed to obtain controlled shape, homogenous narrow size distribution, and enhanced plasmonic properties by tuning the irradiation power. The proposed experimental procedure allows to show the different phenomena involved in the ultrasound assisted Turkevich-Frens method, associating the physical effects such as temperature and pH rise over reaction with cavitation phenomena and radical exchange. The obtained results show the strong influence of increasing ultrasound irradiation on particle size, polyhedral structure and optical properties of as obtained GNPs and the reaction kinetics for their formation. Particle sizes from 16 nm reached in 60 and 150 samples can be reduced to 12 nm using 210 sample procedure. The nucleation, growth, and saturation as irradiation power increase displayed exotic solid configuration of GNPs: tetrakis hexahedron, triakis icosahedron, and pentakis dodecahedron as HRTEM images reveal. The GNPs formation kinetics monitored by SRP in the UV-Vis spectra reveals the enhanced optical properties and LSRP at reduced time (20 min). The seed concentration is optimized by ultrasound irradiation and lead to monodisperse and localized absorption GNPs manufacture. The disproportionation process of three complexes $AuCl_2^-$ form clusters via aurophilic interaction were suggested in order to explain chemical effects in the synthesis. A simple molecular-based scheme was proposed, from the formation and growth processing of GNPs under the reported reaction conditions. Electrons of $AuCl_2^-$ are rapidly exchanged through the gold surface and after $AuCl_4^-$ and $Cl^-$ are desorbed and a complex tetrachloroaurate is recycled for the two-electron reduction by citrate. This mechanism allows the deposition of new gold atoms on the surface and promotes the growth of seeds. The experiments were designed to explore experimental ultrasound irradiation power influence on the optimization of GNPs systems and the results show remarkable improvement of the obtained materials, positioning to the ultrasound assisted methods as an option for nanomaterials elaboration, modification and optimization in a simple and economic way.

**Acknowledgments**

Authors would like to acknowledge Rodrigo Fernandez Pacheco from Laboratorio de Microscopias Avanzadas (LMA) of INA for the facilities in TEM observations, the use of Servicio General de Apoyo a la Investigación-SAI, Universidad de Zaragoza. J.A. Fuentes-García thanks the Mexican council of science and technology (CONACyT) for financial support through a postdoctoral fellowship #711124.

**References**

[1] L.C. Kennedy, L.R. Bickford, N.A. Lewinski, A.J. Coughlin, Y. Hu, E.S. Day, J.L. West, R.A. Drezek, A New Era for Cancer Treatment: Gold-Nanoparticle- Mediated Thermal Therapies, Small 7 (2011) 169–83. https://doi.org/10.1002/smll.201000134

[2] R.A. Sperling, P.R. Gil, F. Zhang, M. Zanella, W.J. Parak, Biological applications of gold nanoparticles, Chem. Soc. Rev. 37 (2008) 1896–1908. https://doi.org/10.1039/b712170a

[3] J.F. Hainfeld, D.N. Slatkin, H.M. Smilowitz, The use of gold nanoparticles to enhance radiotherapy in mice, Phys. Med. Biol. 49 (2004) N309–N315. https://doi.org/10.1088/0031-9155/49/18/N03

[4] P. Ghosh, G. Han, M. De, C.K. Kim, V.M. Rotello, Gold nanoparticles in delivery applications, Adv. Drug Deliv. Rev. 60 (2008) 1307–1315. https://doi.org/10.1016/j.addr.2008.03.016

[5] K.S. Lee, M.A. El-Sayed, Gold and Silver Nanoparticles in Sensing and Imaging: Sensitivity of Plasmon Response to Size, Shape, and Metal Composition, J. Phys. Chem. B 110 (2006) 19220–19225. https://doi.org/10.1021/jp062536y

[6] T.K. Sau, C.J. Murphy, Room Temperature, High-Yield Synthesis of Multiple Shapes of Gold Nanoparticles in Aqueous Solution, J. Am. Chem. Soc. 126 (2004) 8648–8649. https://doi.org/10.1021/ja047846d

[7] S.D. Perrault, W.C.W. Chan, Synthesis and Surface Modification of Highly Monodispersed, Spherical Gold Nanoparticles of 50-200 nm, J. Am. Chem. Soc. 131 (2009) 17042–17043. https://doi.org/10.1021/ja907069u

[8] J. Turkevich, P. Cooper Stevenson, J. Hillier, A Study of the Nucleation and Growth Processes in the Synthesis of Colloidal Gold, Disc. Farad. Soc. 11 (1951) 55–75. https://doi.org/10.1039/DF9511100055

[9] G. Frens, Controlled Nucleation for the Regulation of Particle Size in Monodisperse Gold Suspensions, Nat. Phys. Sci. 241 (1973) 20–22. https://doi.org/10.1038/physci241020a0

[10] J. Kimling, M. Maier, B. Okenve, V. Kotaidis, H. Ballot, A. Plech, Turkevich Method for Gold Nanoparticle Synthesis Revisited, J. Phys. Chem. B 110 (2006) 15700–15707. https://doi.org/10.1021/jp061667w

[11] M. Wuithschick, A. Birnbaum, S. Witte, M. Sztucki, U. Vainio, N. Pinna, K. Rademann, F. Emmerling, R. Kraehnert, J. Polte, Turkevich in New Robes: Key Questions Answered for the Most Common Gold Nanoparticle Synthesis, ACS Nano 9 (2015) 7052–7071. https://doi.org/10.1021/acsnano.5b01579

[12] NEW NEW NEW

[13] J. Xiao, Limin Qi, Surfactant-assisted, shape-controlled synthesis of gold nanocrystals, Nanoscale 3 (2011) 1383–1396. https://doi.org/10.1039/c0nr00814a

[14] X. Wu, C. Lu, Z. Zhou, G. Yuan, R. Xiong, X. Zhang, Green synthesis and formation mechanism of cellulose nanocrystal-supported gold nanoparticles with enhanced catalytic performance, Environ. Sci.: Nano 1 (2014) 71–79. https://doi.org/10.1039/c3en00066d

[15] T. Elavazhagan, K.D. Arunachalam, Memecylon edule leaf extract mediated green synthesis of silver and gold nanoparticles, Int. J. Nanomedicine. 6 (2011) 1265–1278. https://doi.org/10.2147/IJN.S18347

[16] F.K. Liu, Y.C. Chang, F.H. Ko, T.C. Chu, Microwave rapid heating for the synthesis of gold nanorods, Mater. Lett. 58 (2004) 373–377. https://doi.org/10.1016/S0167-577X(03)00504-4

[17] F. Kim, J.H. Song, P. Yang, Photochemical Synthesis of Gold Nanorods, J. Am. Chem. Soc. 124 (2002) 14316–14317. https://doi.org/10.1021/ja028110o

[18] J.U. Kim, S.H. Cha, K. Shin, J.Y. Jho, J.C. Lee, Synthesis of Gold Nanoparticles from Gold(I)-Alkanethiolate Complexes with Supramolecular Structures through Electron Beam Irradiation in TEM, J. Am. Chem. Soc. 127 (2005) 9962–9963. https://doi.org/10.1021/ja042423x

[19] S. Seino, T. Kinoshita, Y. Otome, T. Nakagawa, K. Okitsu, Y. Mizukoshi, T. Nakayama, T. Sekino, K. Niihara, T.A. Yamamoto, Gamma-ray synthesis of

magnetic nanocarrier composed of gold and magnetic iron oxide, J. Magn. Magn. Mater. 293 (2005) 144–150. https://doi.org/10.1016/j.jmmm.2005.01.054

[20] L.F. de Freitas, G.H.C. Varca, J.G. dos Santos, A.B. Lugão, An Overview of the Synthesis of Gold Nanoparticles Using Radiation Technologies, Nanomaterials 8 (2018) 1–23. https://doi.org/10.3390/nano8110939

[21] Ebrahim Alipanahpour Dil, Mehrorang Ghaedi, Arash Asfaram, Fahimeh Zare, Fatemeh Mehrabi, Fardin Sadeghfar. Comparison between dispersive solid-phase and dispersive liquid–liquid microextraction combined with spectrophotometric determination of malachite green in water samples based on ultrasound-assisted and preconcentration under multi-variable experimental design optimization. Ultrasonics - Sonochemistry 39 (2017) 374–383. https://doi.org/10.1039/c5ra02214b

[22] S. Theerdhala, D. Bahadur, S. Vitta, N. Perkas, Z. Zhong, A. Gedanken, Sonochemical stabilization of ultrafine colloidal biocompatible magnetite nanoparticles using amino acid, l-arginine, for possible bio applications, Ultrason. Sonochem. 17 (2010) 730–737. https://doi.org/10.1016/j.ultsonch.2009.12.007

[23] K.S. Suslick, Sonochemistry, Science 247 (1990) 1439-144. https://doi.org/10.1126/science.247.4949.1439

[24] N.S.M. Yusof, B. Babgi, Y. Alghamdi, M. Aksu, J. Madhavan, M. Ashokkumara, Physical and chemical effects of acoustic cavitation in selected ultrasonic cleaning applications, Ultrason. Sonochem. 29 (2016) 568–576. https://doi.org/ 10.1016/j.ultsonch.2015.06.013

[25] D. Chen, S.K. Sharma, A. Mudhoo. Handbook on Applications of Ultrasound. Sonochemistry for Sustainability, Taylor & Francis Group, LLC. CRC, 2012.

[26] T.J. Mason, D. Peters, Practical Sonochemistry: Power Ultrasound Uses and Applications, Woodhead Publishing, 2002.

[27] K. Vilkhu, R. Mawson, L. Simons, D. Bates, Applications and opportunities for ultrasound assisted extraction in the food industry- A review. Innov. Food Sci. Emerg. Technol. 9 (2008) 161–169. https://doi.org/10.1016/j.ifset.2007.04.014

[28] Ali Akbar Bazrafshana, Mehrorang Ghaedi, Shaaker Hajatib, Reza Naghihad, Arash Asfaram. Synthesis of ZnO-nanorod-based materials for antibacterial, antifungal activities, DNA cleavage and efficient ultrasound-assisted dyes adsorption. Ecotoxicology and Environmental Safety 142 (2017) 330–337.

[29] Arash Asfaram, Mehrorang Ghaedi, Mihir Kumar Purkait. Novel synthesis of nanocomposite for the extraction of Sildenafil Citrate (Viagra) from water and urine samples: Process screening and optimization. Ultrasonics Sonochemistry, Volume 38, September 2017, Pages 463-472.

[30] Hossein Zare Khafri, Mehrorang Ghaedi, Arash Asfaram, Mohammad Safarpoor. Synthesis and characterization of ZnS:Ni-NPs loaded on AC derived from apple tree wood and their applicability for the ultrasound assisted comparative adsorption of cationic dyes based on the experimental design. Ultrasonics Sonochemistry, Volume 38, September 2017, Pages 371-380.

[31] Aharon Gedanken. Using sonochemistry for the fabrication of nanomaterials

[32] Stephan Barcikowski, Anton Plech, Kenneth S. Suslick and Alfred Vogel. Materials synthesis in a bubble. MRS Bulletin Volume 44, Issue 5 (Acoustic Processes in Materials)May 2019, pp. 382-39. https://doi.org/10.1557/mrs.2019.107

[33] T. Fujimoto, S. Terauchi, H. Umehara, I. Kojima, W. Henderson, Sonochemical Preparation of Single-Dispersion Metal Nanoparticles from Metal Salts, Chem. Mater. 13 (2001) 1057–1060. https://doi.org/ 10.1021/cm000910f

[34] J.H. Bang, K.S. Suslick, Applications of Ultrasound to the Synthesis of Nanostructured Materials, Adv. Mater. 22 (2010) 1039–1059. https://doi.org/10.1002/adma.200904093

[35] H. Nomura, S. Koda, What Is Sonochemistry? Sonochemistry and the Acoustic Bubble. Elsevier 2015.

[36] Hinman, J.J., Suslick, K.S. Nanostructured Materials Synthesis Using Ultrasound. Top Curr Chem (Z) 375, 12 (2017). https://doi.org/10.1007/s41061-016-0100-9.

[37] Kumari S.Ojha, Brijesh K.Tiwari, Colm P.O'Donnell. Chapter Six - Effect of Ultrasound Technology on Food and Nutritional Quality. Advances in Food and Nutrition Research Volume 84, 2018, Pages 207-240. https://doi.org/10.1016/bs.afnr.2018.01.001

[38] K. Okitsu, M. Ashokkumar, F. Grieser, Sonochemical Synthesis of Gold Nanoparticles: Effects of Ultrasound Frequency, J. Phys. Chem. B 109 (2005) 20673–20675. https://doi.org/10.1021/jp0549374

[39] J.E. Park, M. Atobe, T. Fuchigami, Synthesis of multiple shapes of gold nanoparticles with controlled sizes in aqueous solution using ultrasound, Ultrason. Sonochem. 13 (2006) 237–241. https://doi.org/10.1016/j.ultsonch.2005.04.003

[40] Md Hujjatul Islam, Michael T.Y.Paul, Odne S.Burheim, Bruno G. Pollet. Recent developments in the sonoelectrochemical synthesis of nanomaterials. Ultrasonics Sonochemistry Volume 59, December 2019, 104711. https://doi.org/10.1016/j.ultsonch.2019.104711

[41] Narges Elahi, Mehdi Kamali, Mohammad Hadi Baghersad. Recent biomedical applications of gold nanoparticles: A review. Talanta Volume 184, 1 July 2018, Pages 537-556. https://doi.org/10.1016/j.talanta.2018.02.088

[42] Lei Qin, Guangming Zeng, Cui Lai, Danlian Huang, Piao Xua, Chen Zhang, Min Cheng, Xigui Liu, Shiyu Liu, Bisheng Li, Huan Yi. "Gold rush" in modern science: Fabrication strategies and typical advanced applications of gold nanoparticles in sensing. Coordination Chemistry Reviews Volume 359, 15 March 2018, Pages 1-31. https://doi.org/10.1016/j.ccr.2018.01.006

[43] Jaber Beik, Maziar Khateri, Zohreh Khosravi, S. Kamran Kamrava, Siavash Kooranifar, Habib Ghaznavi, Ali Shakeri-Zadeh. Gold nanoparticles in combinatorial cancer therapy strategies. Coordination Chemistry Reviews Volume 387, 15 May 2019, Pages 299-324. https://doi.org/10.1016/j.ccr.2019.02.025

[44] S. Ali, V.S. Myasnichenko, E.C. Neyts, Size-dependent strain and surface energies of gold nanoclusters, Phys. Chem. Chem. Phys. 18 (2016) 792–780. https://doi.org/10.1039/c5cp06153a

[45] A.Q. Zhang, D.J. Qian, M. Chen, Simulated optical properties of noble metallic nanopolyhedra with different shapes and structures, Eur. Phys. J. D 67 (2013) 1–9. https://doi.org/10.1140/epjd/e2013-40240-1

[46] D.Y. Kim, S.H. Im, O.O. Park, Y.T. Lim, Evolution of gold nanoparticles through Catalan, Archimedean, and Platonic solids, Cryst. Eng. Comm. 12 (2010) 116–121. https://doi.org/10.1039/b914353j

[47] Gong, J., Newman, R., Engel, M. et al. Shape-dependent ordering of gold nanocrystals into large-scale superlattices. Nat Commun 8, 14038 (2017). https://doi.org/10.1038/ncomms14038

[48] P.F. Damasceno, M. Engel, S.C. Glotzer, Predictive self-assembly of polyhedra into complex structures, Science 337 (2012) 453-457. https://doi.org/10.1126/science.1220869

[49] M. Wuithschick, S. Witte, F. Kettemann, K. Rademann, J. Polte, Illustrating the formation of metal nanoparticles with a growth concept based on colloidal stability, Phys. Chem. Chem. Phys. 17 (2015) 19895–19900. https://doi.org/10.1039/c5cp02219c

[50] K. Takiyama, Formation and Aging of Precipitates. VIII.* Formation of Monodisperse Particles (1) Gold Sol Particles by Sodium Citrate Method, Bull. Chem. Soc. Jpn. 31 (1958) 944–950. https://doi.org/10.1246/bcsj.31.944

[51] Minh Tran, Rebekah DePenning, Madeline Turner and Sonal Padalkar. Effect of citrate ratio and temperature on gold nanoparticle size and morphology. Mater. Res. Express 3 (2016) 105027. https://doi:10.1088/2053-1591/3/10/105027

[52] Wenchao Ding, Peina Zhang, Yijing Li, Haibing Xia, Dayang Wang, Xutang Tao. Effect of Latent Heat in Boiling Water on the Synthesis of Gold Nanoparticles of Different Sizes by using the Turkevich Method. ChemPhysChem. 16 (2) (2015) 447-454 https://doi.org/10.1002/cphc.201402648

[53] R. Jin, C. Zeng, M. Zhou, Y. Chen, Atomically Precise Colloidal Metal Nanoclusters and Nanoparticles: Fundamentals and Opportunities, Chem. Rev. 116 (2016) 10346–10413. https://doi.org/10.1021/acs.chemrev.5b00703

[54] Ali Demirci Michael Ngadi Microbial Decontamination in the Food Industry 1st Edition. Novel Methods and Applications Woodhead Publishing 2012.

[55] S. Biggs, P. Mulvaney, C. F. Zukoski, and F. Grieser. Study of Anion Adsorption at the Gold-Aqueous Solution Interface by Atomic Force Microscopy. J. Am. Chem. SOC. 1994,116, 9150-9157.

[56] Lihua Pei, Koichi Mori, and Motonari Adachi. Formation Process of Two-Dimensional Networked Gold Nanowires by Citrate Reduction of AuCl4 - and the Shape Stabilization. Langmuir 2004, 20, 7837-7843.

[57] Jennifer F. Wall, Franz Grieser and Charles F. Zukoski. Monitoring chemical reactions at the gold/solution interface using atomic force microscopy

[58] Marc Breitbach, Dieter Bathen, and Henner Schmidt-Traub. Effect of Ultrasound on Adsorption and Desorption Processes. Ind. Eng. Chem. Res. 2003, 42, 5635-5646.

[59] I. Ojea-Jiménez, J.M. Campanera, Molecular Modeling of the Reduction Mechanism in the Citrate-Mediated Synthesis of Gold Nanoparticles, J. Phys. Chem. C 116 (2012) 23682−23691. https://doi.org/10.1021/jp305830p

[60] I. Ojea-Jiménez, N.G. Bastús, V. Puntes, Influence of the Reagents Addition in the Citrate-Mediated Synthesis of Gold Nanoparticles, J. Phys. Chem. C 115 (2011) 15752−15757. https://doi.org/10.1021/jp2017242

[61] J.L. Wang, L.J. Xu, Advanced Oxidation Processes for Wastewater Treatment: Formation of Hydroxyl Radical and Application, Crit. Rev. Environ. Sci. Technol. 42 (2012) 251–325. https://doi.org/10.1080/10643389.2010.507698

[62] M.C. Gimeno, The Gold Chemistry. In Modern Supramolecular Gold Chemistry: Gold-Metal Interactions and Applications. A. Laguna, Wiley-VCH Verlag GmbH & Co. KGaA, 2008.

[63] E. Janata, A. Henglein, B.G. Ershov, First Clusters of Ag+ Ion Reduction in Aqueous Solution, J. Phys. Chem. 98 (1994) 10888–10890. https://doi.org/10.1021/j100093a033

[64] D.W. Larson, M.W. Lister, Catalytic decomposition of acetonedicarboxylic acid, Can. J. Chem. 46, (1968) 823–832. https://doi.org/ 10.1139/v68-143

[65] C.H. Gammons, Y. Yu, A.E. Williams-Jones, The disproportionation of gold(I) chloride complexes at 25 to 200°C, Geochim. Cosmochim. Acta 61 (1997) 1971–1983. https://doi.org/ 10.1016/S0016-7037(97)00060-4

[66] K. Theilacker, H.B. Schlegel, M. Kaupp, P. Schwerdtfeger, Relativistic and Solvation Effects on the Stability of Gold(III) Halides in Aqueous Solution, Inorg. Chem. 54 (2015) 9869–9875. https://doi.org/ 10.1021/acs.inorgchem.5b01632

[67] G.V. Buxton, C.L. Greenstock, W.P. Helman, A.B. Ross, Critical Review of Rate Constants for Reactions of Hydrated Electrons, Hydrogen atoms and Hydroxyl Radicals (•OH/•O-), J. Phys. Che. Ref. Data 17 (1988) 513-886. https://doi.org//10.1063/1.555805

[68] A. Sarkar, E. Janata, Formation of the Silver Hydride Ion AgH+ upon the Reduction of Silver Ions by H• in Aqueous Solution. A Pulse Radiolysis Study, Z. Phys. Chem. 221 (2007) 403–413. https://doi.org/10.1524/zpch.2007.221.3.403

[69] X. Liu, H. Xu, H. Xia, D. Wang, Rapid Seeded Growth of Monodisperse, Quasi-Spherical, Citrate-Stabilized Gold Nanoparticles via H2O2 Reduction, Langmuir 28 (2012) 13720–13726. https://doi.org/10.1021/la3027804

[70] M.H. Enterazi, P. Kruus, Effect of frequency on sonochemical reactions II. Temperature and intensity effects, Ultrason. Sonochem. 3 (1996) 19–24. https://doi.org/10.1016/1350-4177(95)00037-2

[71] V. Augugliario, M. Bellardita, V. Loddo, G. Palmisano, L. Palmisano, S. Yurdakal, Overview on oxidation mechanisms of organic compounds by TiO2 in heterogeneous photocatalysis, J. Photochem. Photobiol. C Photochem. Rev. 13 (2012) 224–245. https://doi.org/ 10.1016/j.jphotochemrev.2012.04.003

[72] M. Doyen; K. Bartik, G. Bruylants, UV-Vis and NMR study of the formation of gold nanoparticles by citrate reduction: Observation of gold-citrate aggregates, J. Colloid Interface Sci. 399 (2013) 1–5. https://doi.org/ 10.1016/j.jcis.2013.02.040

[73] M. Álvaro, C. Aprile, A. Corma, B. Ferrer, H. García, Influence of radical initiator in gold catalysis: Evidence supporting trapping of radicals derived from azobis(butyronitrile) by gold halides, J. Catal. 245 (2007) 249–252. https://doi.org/10.1016/j.jcat.2006.10.003